\documentclass[10pt,journal]{IEEEtran}

\usepackage{subfigure}
\usepackage{graphicx}
\usepackage{multirow}
\usepackage{enumitem}
\usepackage{caption} 
\usepackage{xcolor}
\usepackage{comment}
\usepackage{listings}

\ifCLASSOPTIONcompsoc
  \usepackage[nocompress]{cite}
\else
  \usepackage{cite}
\fi

\ifCLASSOPTIONcompsoc
  \usepackage[caption=false,font=normalsize,labelfont=sf,textfont=sf]{subfig}
\else
  \usepackage[caption=false,font=footnotesize]{subfig}
\fi

\begin{document}
\title{Evolution of Android's Permission-based Security Model and Challenges}
%

\author{Rajendra Kumar Solanki and Vijay Laxmi and Manoj Singh Gaur~\IEEEmembership{Senior Member,~IEEE}}%

\markboth{IEEE COMMUNICATIONS SURVEYS AND TUTORIALS,~VOL.~00, NO.~0, April~2021}%
{Rajendra \MakeLowercase{\textit{et al.}}: Evolution of Android's Permission-based Security Model and Challenges}



\maketitle

\begin{abstract}
Android Permission Model and Application (app) analysis has consistently remained the focus of the investigation of research groups and stakeholders of the Android ecosystem since it was launched in 2008. Even though the Android smartphone operating system (OS) permission model has evolved significantly from `all-or-none access' to `user-chosen dangerous resource access', specific challenges and issues remain unresolved even after 15 years after the smartphone OS launch. This study addresses the issues and documents the research work in this arena through a comprehensive literature survey and comparative analysis.  

The survey's focal point is the Android permission model and relevant research between 2010-2022. We systematize the knowledge on (i) Android API Calls to permissions mapping, (ii) Android Permissions evolution, and (iii) how permissions are checked. Furthermore, the survey identifies the permission-related issues and relevant research addressed during the last decade. We reference seminal work in these areas. We summarize the identified research gaps and present future directions for early and experienced researchers. 
\end{abstract}


%
\IEEEpeerreviewmaketitle


\section{Introduction}\label{sec:introduction}
\subsection{Motivation}
Smartphone apps download and usage statistics \cite{URL:buidfire2021} reveal an interesting trend where the users spend most of their smartphone time on third-party developer apps. The average smartphone user avails ten apps per day and approximately thirty apps every month. We use apps for accessing social presence, reading news, online banking, e-commerce, productivity improvement, games, etc. The smartphone app-based economy has exceeded all expectations and proliferated in the past decade. Over two and a half million Android apps run on billions of Android-powered devices \cite{URL:statistaApps2022},  \cite{URL:StatistaAndyPlatforms}. 

With the growth of smartphone users across the globe, Android has gained a significant operating system market share by the number of devices being shipped \cite{URL:idc2020}, and it is one of the most widely deployed operating systems. This trend continues to grow further. By the year 2023, the smartphone economy is expected to garner revenue of over \$935 billion \cite{URL:buidfire2021}. 
Such a vast revenue-generating platform has also attracted malware writers who steal personal data for monetary gains and to track users based on their activities on Android devices. 

Privacy of Personally Identifiable Information (PII) of users is at risk when users are unaware of malicious and unwanted software disguised in the mobile apps \cite{pii_sigcomm10.1145/1672308.1672328}, \cite{pii_bestpractices}. End users submit a lot of sensitive data on the apps. Apps can ask for far more permissions than actually required. This behaviour raises privacy and security concerns for such deceptive apps \cite{rajendra2019mapper}, \cite{SOLANKI2021102493}, \cite{google_privacy-deception}.

In the Android ecosystem, the responsibility to ensure the security and privacy of data lies with the Android system, app developers and end-users \cite{10.1145/2811411.2811493}. Given the potentially diverse user base and malware reported in the past decade, the user layer appears a weak link to executing decisions related to security and privacy. 
Apps and associated unwanted side effects impact our daily lives. A social networking site's choices can influence democracy \cite{URL:CambridgeAnalyticaVox2018}, \cite{URL:CambridgeAnalyticaGuardian2018}. Video conferencing apps can monitor microphones and activities even when they are muted \cite{https://doi.org/10.48550/arxiv.2204.06128}. These decisions cannot be left to users alone that they will figure it all out. 

Hence, the Android system must be equipped with reasonable measures toward a strong security policy, and app developers should enforce the implementation. Third-party app developers can enforce such policy decisions provided they are equipped with the knowledge of capabilities corresponding to the appropriate Android API calls. 

To enforce the protection of system resources and access control, Android uses permissions. Each of these permissions governs access to resources through use of API calls. Developers should be made aware of how a specific permission maps to related API calls. 
Here, a precise mapping between Android API calls and the corresponding Android permissions is of great significance. The mapping can be utilized to enforce security policy for malware detection frameworks \cite{arp2014drebin}, for deceptive app detection frameworks and studies of over-privileged application binaries \cite{SOLANKI2021102493}, \cite{10.1016/j.cose.2014.10.005}, \cite{Stowaway10.1145/2046707.2046779}, \cite{Pscout10.1145/2382196.2382222}. 

Despite the impact of API mapping data to research, lately, there is little work on how Android APIs map to permissions in practice apart from a few limited or now obsolete studies \cite{Stowaway10.1145/2046707.2046779},\cite{Pscout10.1145/2382196.2382222},\cite{axplorer197193}, \cite{DyperminLYVAS2018472}. These papers have impacted thousands of studies by looking at combined citations as on early 2022 that show the impact and significance of the permission specifications-related studies. 
In this paper, we present a comprehensive analysis of the use of Android permissions in Android research publications. 

Another aspect we explore is to study the evolution of the Android ecosystem from a permissions point of view. In this direction, significant milestones are the introduction of runtime permissions, revoking permissions post-installation, automatically granted permissions, one-time permissions, permission validity during usage only, and restricting specific permissions from apps other than default handler, etc. One more aspect is of interest as to how runtime permissions are managed actually - checking grant status, requesting and explaining permissions, and finally storing results of the user's response to permissions prompts. Not requesting and not checking granted permissions can result in Android Runtime Permission (ARP) issues \cite{ARPissues9705152}. 

We begin with a study of the most extensive work that dissects the Android Permission specification from Android source code and techniques used over the past decade. Further, we show the evolution of the permission model and how permissions are checked. We present research work that has happened around these topics. 

\textbf{Contributions}: The major contributions of this study are aligned with investigating the following research questions:
\begin{enumerate}[label=\alph*)]

  \item \textbf{RQ1: What contributions have been made for generating a mapping between Android API calls and permissions required?} With this question, we survey the various research works carried out to provide a mapping between API and permissions, their relevance, and applicability today for modern apps. We also investigate whether the proposed tools and datasets are available to reproduce the results in Section \ref{sec:mapping_between_API_calls_and_permissions}. 

  \item \textbf{RQ2: How Android ecosystem has evolved with regard to the permission-based model?} With the growth and release of several Android API levels over the past decade, we survey the evolution of the permission-based model for Android users, developers, and how the research work has advanced on this in Section \ref{sec:evolution_permission_model}. 
  \item \textbf{RQ3. How permissions are checked?} With a lot of literature available on Android, at times it is often not clear how permissions-based protection is enforced, we address this through new lenses in Section \ref{sec:checking_permissions}. 

  \item \textbf{RQ4. What are the research gaps that should be addressed?} Finally, we survey the challenges that are yet to be benefited from the research work of a decade. 

\end{enumerate}

\subsection{Publications and datasets extraction}
To find proposed mechanisms, tools, and datasets of the publications to be considered we leverage well-known electronic repositories and publication indices, namely IEEE Xplore, ACM Digital Library, ScienceDirect, SpringerLink, and DBLP. This survey considers publications from some of the reputed conferences and journals like IEEE Transactions on Information Forensics and Security (TIFS), Computers and Security, IEEE Communications Survey and Tutorials, USENIX Security, ACM SIGSAC Conference on Computer and Communications Security (CCS), ACM Computing Surveys (CSUR), etc. for the period 2011-2020. This survey restricts the focus to permissions and privacy concerns while looking into the publications in (i) Security, (ii) Privacy, and (ii) Software Engineering domains. We follow some of the best practices for the survey on security and privacy research \cite{Redmiles2017ASO}. 

\begin{table*}
\renewcommand{\arraystretch}{1.8}
\caption{Comparison of recent surveys on Android and Permissions}
\label{tab:comparison_other_surveys}
\centering
\begin{tabular}{ c | c | c | c | c | c | c }
\hline
\bfseries Related Study & \bfseries Year & \bfseries Permission & \bfseries Evolution of & \bfseries Permission & \bfseries Malware & \bfseries Discussion Post \\ 
 &  & \bfseries Map & \bfseries Permissions & \bfseries Checks & \bfseries Detection & \bfseries API level 23 \\ \hline
This study & 2022 & yes & yes & yes &  & yes \\ \hline
Mayrhofer et al. \cite{rene10.1145/3448609} & 2021 &  & yes &  &  &  \\ \hline
Iman et al. \cite{ImanAala9272963} & 2020 &  & yes &  &  & yes \\ \hline
Bakour et al. \cite{BakourSN10.1007/s42452-019-1124-x} & 2019 &  &  &  & yes & \\ \hline
Tam et al. \cite{TamCSUR1710.1145/3017427} & 2017 &  &  &  & yes & \\ \hline
Faruki et al. \cite{ParvezCOMST6999911} & 2016 &  &  &  & yes & \\ \hline 
Sufatrio et al. \cite{sufatrio10.1145/2733306} & 2015 &  & yes &  & yes &  \\ \hline 
\end{tabular}
\end{table*}

Table \ref{tab:comparison_other_surveys} shows a comparison on how this study is different from surveys published in the recent past. No other study has addressed research work on permission mapping, permission checks, and evolution of permissions including coverage on post API levels 23 when runtime permissions were added. Several studies have discussed malware detection frameworks \cite{BakourSN10.1007/s42452-019-1124-x}, \cite{TamCSUR1710.1145/3017427}, \cite{ParvezCOMST6999911}, \cite{sufatrio10.1145/2733306} and a few have defined security model \cite{rene10.1145/3448609}, \cite{TamCSUR1710.1145/3017427}.

\textbf{Organization}: The rest of the paper is organized as follows: %
Section \ref{sec:background} provides the essential background information for the study; specifically, we explore the Android operating system, security model, and its challenges. %
Section \ref{sec:mapping_between_API_calls_and_permissions} covers the studies of providing a precise mapping between Android API calls and permissions. %
Section \ref{sec:evolution_permission_model} presents the major evolutionary work around Android's permission-based security model and relevant research efforts in this direction. %
Section \ref{sec:checking_permissions} investigates how Android permissions are checked in Java and the native code of the framework. %
We present the lessons learned from the literature and future directions in Section \ref{subsec:reasons_of_permission_gap} and \ref{sec:future_directions} as main takeaways. %
Finally, Section \ref{sec:conclusion} concludes this study.

\begin{figure}
    \centering
    {
        \subfigure
        {
            \includegraphics[width=90mm]{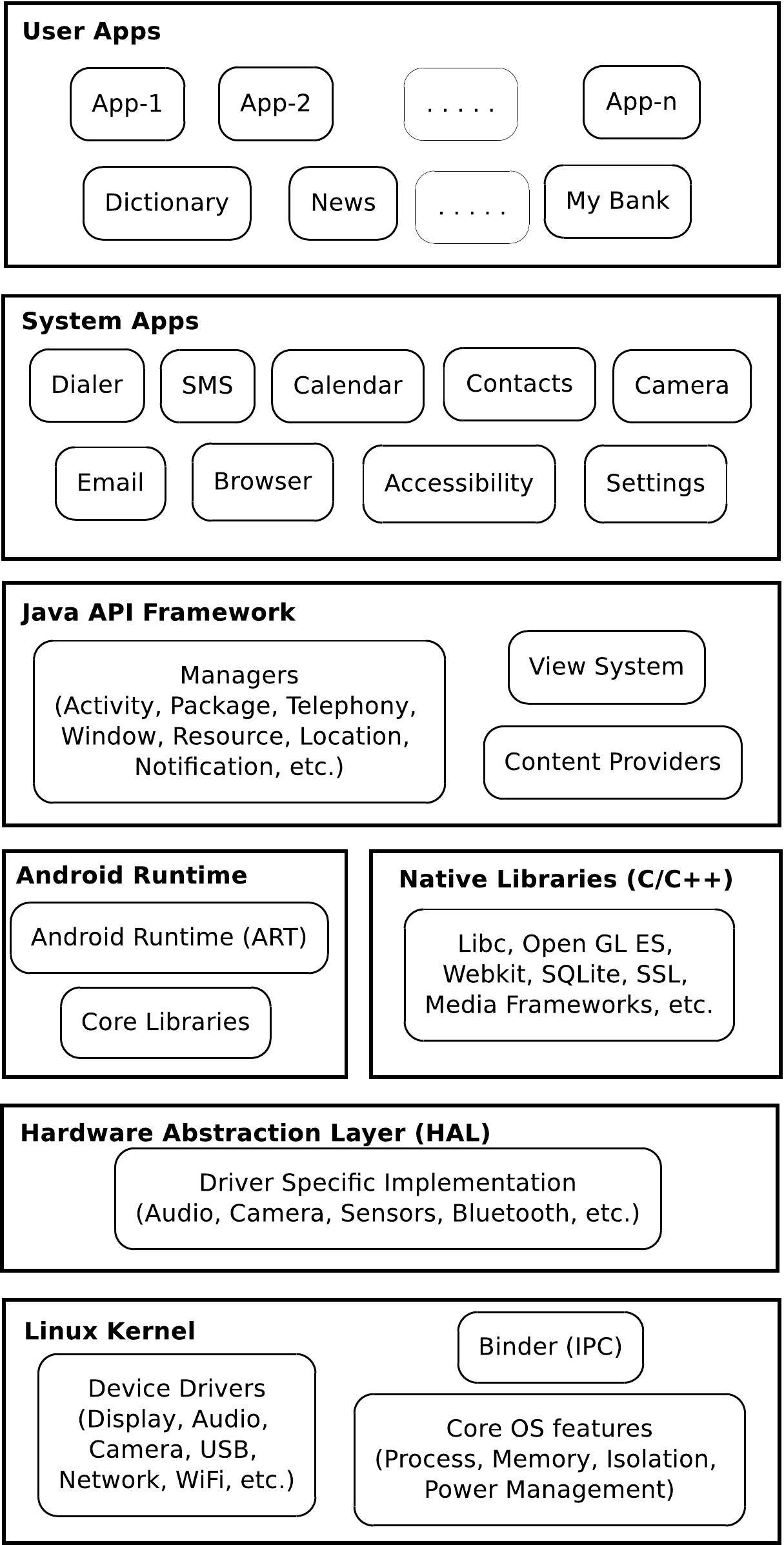}
        }
    }
    \caption{Android Architecture}
    \label{fig:android_architecture}
\end{figure}

\section{Background information on Android}\label{sec:background}
In this background section, we first study the Android application in section \ref{subsec:android_app}. This section reviews and explains the key terminology to aid the readers of this study.

\subsection{Android Architecture}\label{sec:android-architecture}
Android is an open-source, Linux kernel-based operating system (OS) created for modern devices that offers an app-like experience and a rich user interface. Android OS is used in several types of devices including smartphones, tablets, smart displays, televisions, voice-based digital assistants, etc. Android framework consists of six major layers as depicted in Fig - \ref{fig:android_architecture}. The layers are presented from the bottom to the top in the stack as follows: (i) Linux kernel, (ii) Hardware Abstraction Layer (HAL), (iii) Native libraries (C/C++) and Android Run-time (ART), (iv) Java API Framework, (v) System Apps layer, and (vi) User Apps layer. 

The modified Linux kernel contains device drivers, Binder for inter-process communication, and core OS features, including permission-based protection. HAL contains driver-specific implementation for OEMs \cite{android-arch}. The Java API Framework consists of different APIs to create Android apps in the User Apps layer. This API includes managers for various services, View System, and content providers. The API simplifies the reuse of core components. The System Apps layer consists of core apps for a phone call, SMS, email, calendar, contacts, web browsing, camera, and many more resources. The User Apps layer consists of the Android apps pre-installed on device or the apps that can be downloaded by users from the app stores. 

\subsubsection{Android Security}
From a system security perspective, the primary goals of the Android operating system are to protect system resources, protect user data and provide isolation for apps. Towards achieving these goals, the following security mechanisms are provided by Android OS. 
\begin{itemize}
    \item \textbf{Linux Kernel}: provides the core security features, user and group-oriented access control (file permissions) using Linux UID/GID, process isolation, secure inter-process communication, SE Linux policy enforcement, etc. 
    %
    \item \textbf{Mandatory Application Sandbox}: for each process. Two separate apps will run under two different UIDs. 
    \item \textbf{Application signing}: Each app that runs on the Android device must be signed by app developers using a self-signed certificate or a 3rd-party OEM. App signing allows app developers to identify the authors or publishers of the app. The signed app certificate defines the user ID associated with the app and facilitates application sandboxing \cite{app_signing}. 
    \item \textbf{Application level permissions}: each app declares permissions to access system hardware or specific functionalities, and subsequently they are consented to or denied by the user. More details are presented in the sections below. 
\end{itemize}

\subsection{Android Application}\label{subsec:android_app}
\begin{figure}
    \centering
    {
        \subfigure
        {
            \includegraphics[width=90mm]{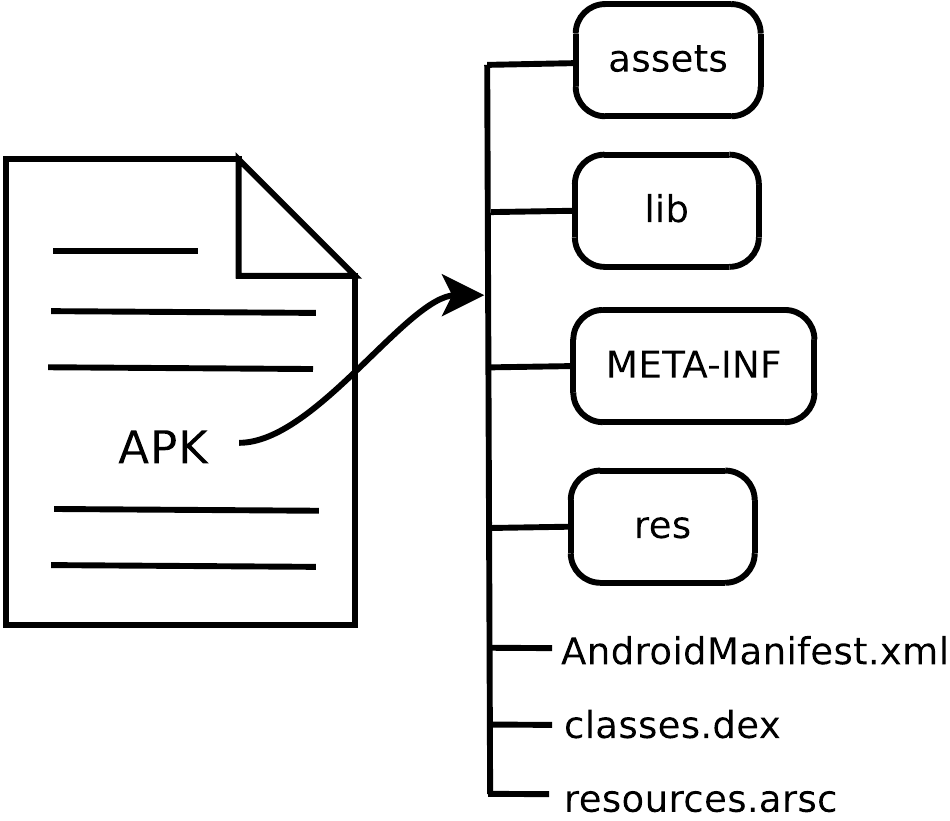}
        }
    }
    \caption{Android application}
    \label{fig:android_application}
\end{figure}

Android application or app is stored on the device in a compressed zip-aligned file - Application PacKage (APK) binary. An APK file contains files such as AndroidManifest.xml, classes.dex, resources.arsc) and folders such as assets, lib, META-INF, res, etc.) as shown in Figure \ref{fig:android_application}. 
The AndroidManifest.xml file contains the app deployment layout, permissions, Intents, and content provider details. The classes.dex file contains Java bytecode - including classes, and interface implementation - packaged in a single .dex file. The resources.arsc file is an archive of compiled and un-compiled resources. The compiled resources consist of the string resource XML files and layout resource XML files typically stored in the res/xml subdirectory. The un-compiled resources consist of raw files such as images, audio, and video files typically stored in the res/raw subdirectory. 

\emph{Android app runtime environment}: Android apps require a virtual machine to run, which is called Dalvik Virtual Machine (DVM) prior to API level 5.0 or Android Run Time (ART) from API level 5.0 onward. Both ART and Dalvik runtime execute Dalvik Executable (Dex) bytecode format. ART supports multiple Dex files in APK binary natively. ART uses Ahead Of Time (AOT) compilation to perform pre-compilation at app installation time by scanning one or more Dex files and compiling them into a single .oat file for execution by the Android device in order to improve performance \cite{multi-dex}.

\begin{figure*}
    \centering
        \includegraphics[width=160mm,height=100mm]{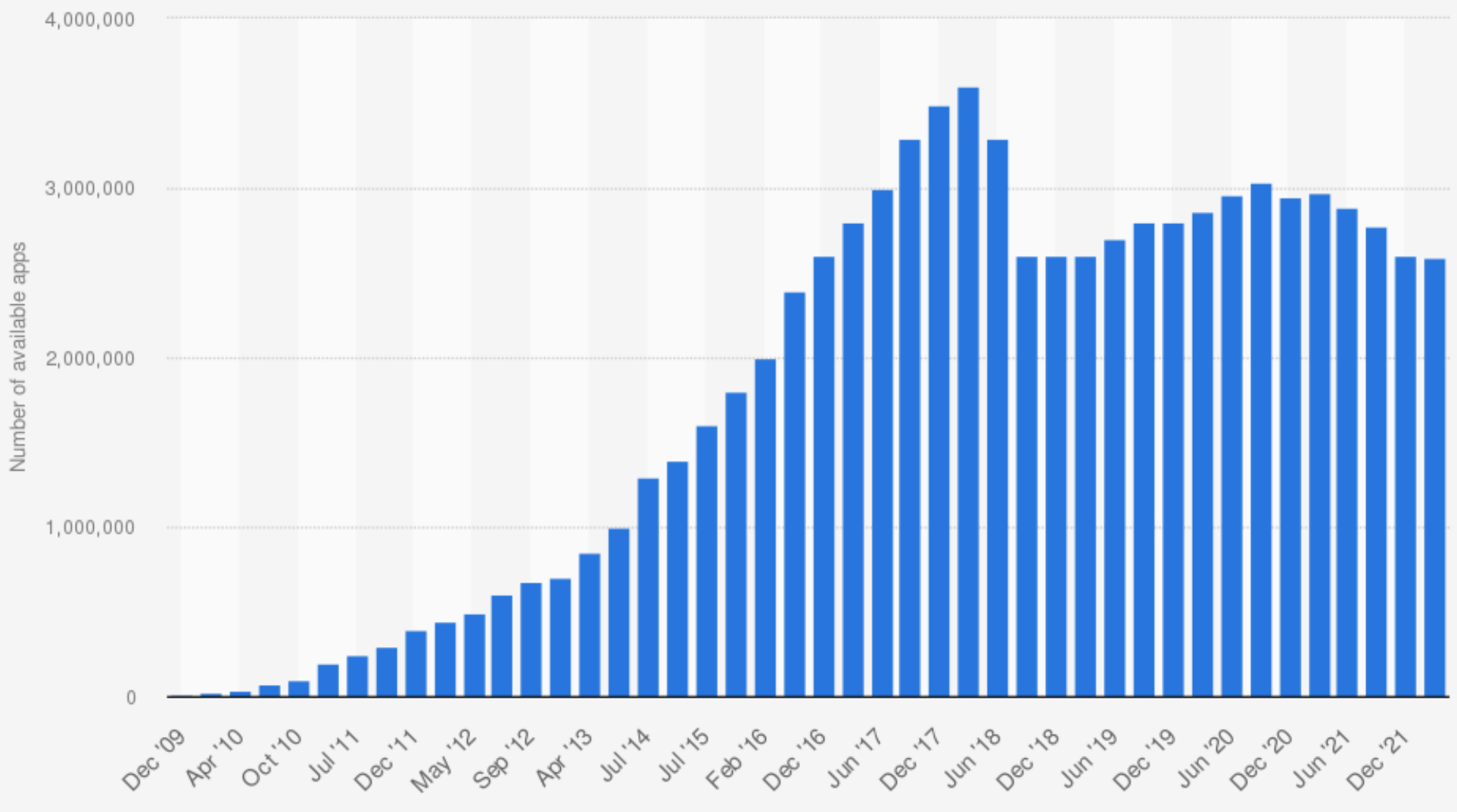}
    \caption{Number of available apps in Google Play Store during 2009-2022}
    \label{fig:gps_apps_2009-22}
\end{figure*}

\emph{Availability of apps}: There are some apps pre-installed on Android devices and users can install apps from Google Play Store. The official App repository for Android is known as Google Play. It was earlier known as Android Market when launched in 2008. Android allows apps with a valid developer-signed certificate. There are alternative app markets available, such as Samsung Galaxy Apps, F-Droid, Amazon App Store, Huawei App Store, etc. Figure \ref{fig:gps_apps_2009-22} shows the number of Android apps available on the Google Play Store between the first quarter of 2015 to 2021. The first time, the number of available apps declined in June 2018 when Google Play Store removed close to a million unwanted apps \cite{gps_removed_apps}, \cite{AA_gps_purge}, \cite{URL:playprotect}. 

An Android application contains the following components: 

\begin{enumerate}[label=\alph*)]
    \item Activities: these components are the screens or user interface (UI) components that the end-user interacts with. One activity can communicate with other activities using intents. 
    \item Services: these application components run in the background. These are like daemons running to serve other running processes. 
    \item Broadcast Intents and Receivers: these components send messages that all other applications or certain individual applications receive. 
    \item Content Providers: these components are data stores of applications and can share data between applications. 
\end{enumerate}
%

\begin{figure*}
    \centering
    {
        \subfigure
        {
            \includegraphics[width=160mm,height=120mm]{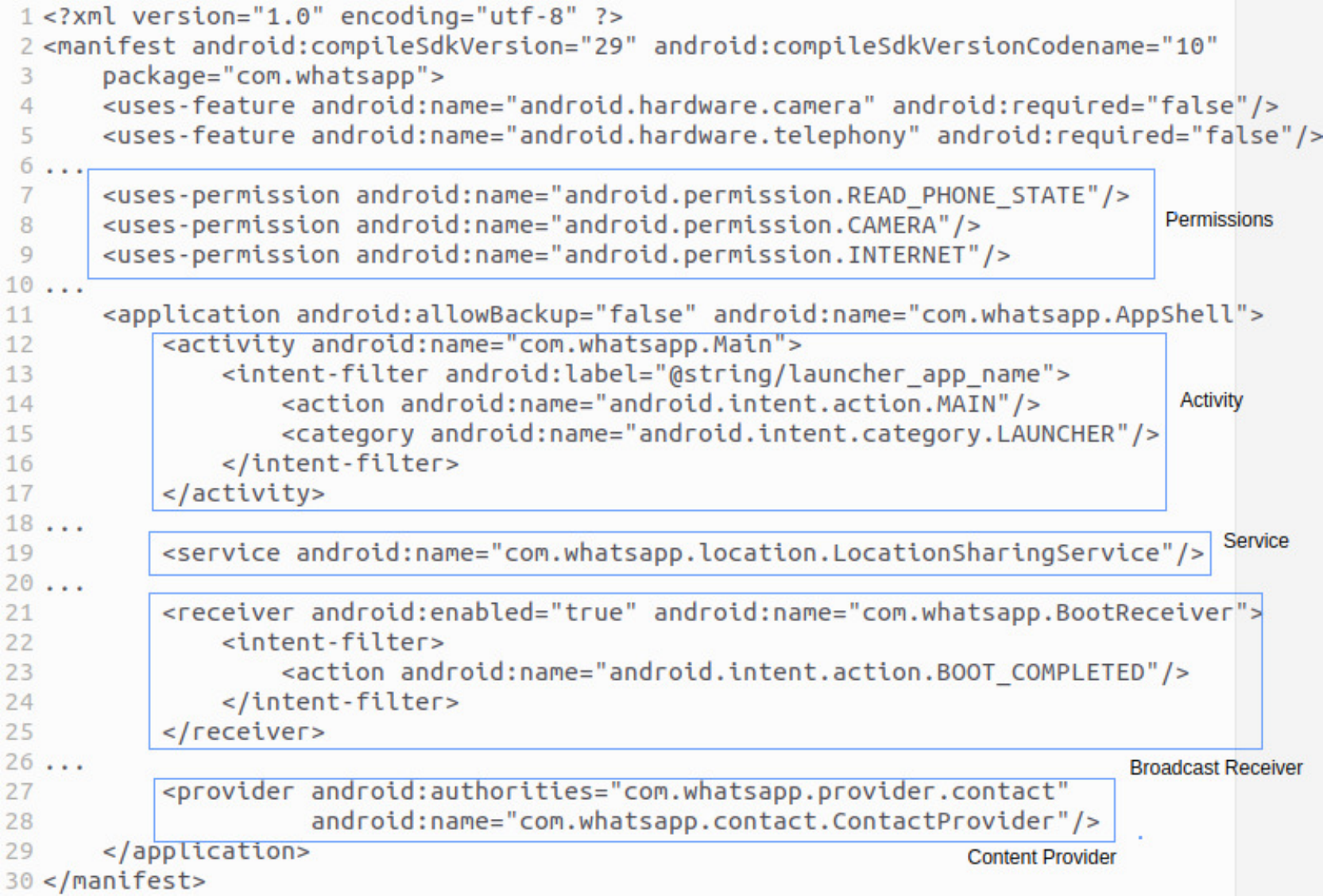}
        }
    }
    \caption{App manifest showing Activity, Service, Broadcast Receiver, and Content Provider}
    \label{fig:sample_app_manifest}
\end{figure*}

Android added Play Protect \cite{URL:playprotect} to detect Potentially Harmful Applications (PHAs) installed or to be installed on the device. Google Play Protect is a real-time malware scanner to scan all applications present on the device irrespective of the source they were installed from. Android Security Report published in 2018, revealed that 0.08\% of devices using Google Play alone for downloading apps were affected by PHAs \cite{URL:GoogleAndySecReport2018}. Looking at the install base, still this can be a fairly large number of devices around. 

\subsection{Attack Surface}\label{subsec:app_weaknesses}
In this section, we review the attack surface for Android applications and how malware writers can exploit this potentially to circumvent permission-based security mechanisms. In the context of Android apps, reducing the attack surface is often associated with aligning to the principle of least privileges. 

Prominent entry points exploited by Android malware or deceptive apps are as follows:

\begin{itemize}
    \item Sensitive Resources: User's personal information and activities are prone to exposure through a phone call, SMS, Contacts, Calendars, Location, and sensors. 
    
    \item Device Tracking Identifiers: Android app developers can track users or devices in near real-time using unique device identifiers such as IMEI, SIM Subscriber ID, SIM Card Serial number, Hardware Serial number, etc. Using this identifiers, user's information stored on the device can be linked to these identifiers and thus leading to tracking information per user level \cite{track-razaghpanah2018}. Many times, this can be a result of misconfigured or over-privileged app that has obtained a permission READ\_PHONE\_STATE - dangerous permission belonging to permission group Phone. 

    \item User Profile Identifiers: Android Device ID and Google Service Framework (GSF) ID are used by advertisers and other potential tracker sites to exfiltrate information on per app and per-user basis. If access to these IDs is not protected appropriately, this can pose risk to users' privacy. 
    
    \item User or Device Location: Deceptive apps can obtain location data of the user or device using network or WiFi access-related normal permissions or location-related dangerous permissions. The exact location coordinates of a device and user can be tracked by apps using WiFi networks or cellular towers it is connected to. Therefore the location tracking features should be restricted to apps whose core function is to provide location details like GPS trackers with the explicit consent of the user. 

    \item Hardware of Android device: Apps can record audio or video or take pictures without the knowledge of the user. Apps can record and transmit sensor data related to health and user activities. Apps can monitor other processes running in the background and can aid in tracking users based on the same device in use. 
    
    \item Multiple active Android API levels: Since 2014, there have been seven or more Android API levels active at the same time \cite{URL:StatistaAndyPlatforms}. This increases the attack surface of Android apps because despite of new OS releases and improved security and privacy controls being available, apps do not benefit from them. There is no incentive for device manufacturers to upgrade OS major releases faster, thereby leaving no choice for app developers to support prolonged backward compatibility. 
    
\end{itemize}

\subsection{Concepts related to study}\label{subsec:challenges} 
We present the important concepts referenced in this study and those relevant for Android research in particular. 

\subsubsection{Principle of Least Privileges (PoLP)}
Every Android app and referenced library program should operate using the least set of privileges necessary to complete the functionality of the app. This principle shows the significance of the damage that can result from an accident or error characterized by escalation of privileges (EoP). The principle of least privileges emphasizes reducing the number of possible interactions among privileged binaries to the minimum for expected functionality so that unintentional, unwanted, or improper usages of privilege are less likely to occur in any situation. Therefore, if one needs to figure out misuse of a privilege, the number of programs and their interactions that must be checked is minimized \cite{Saltzer1451869}. When this study uses the term "over-privilege", this refers to the violation of the principle of least privileges. 

\subsubsection{Contextual Integrity}
Nissenbaum's theory of "Privacy as Contextual Integrity" emphasizes that information collection and dissemination (or flow), should respect contextual norms of information flow \cite{NissenbaumCI}. Further, privacy violation occurs when either contextual norms of appropriateness or norms of information flow have been violated. By circumventing the permission-based model (automatically granting permissions, tracking persistent identifiers, etc.), apps can ex-filtrate personal data and distribute it outside the device against the expectations of users. If apps are not explicitly requesting and obtaining access to users' personal information, then the association of the user's data with the app or with any 3rd-party violates contextual integrity of the user's data and thus privacy \cite{MapperDroidSOLANKI2021102493}.  

\subsubsection{Notice and Consent Framework (NCF)}
The "Notice and Consent" framework governs privacy practices: organizations can notify users about their privacy policies and practices, and users can consent to these practices explicitly. Notice and Consent to accept cookies on websites and permissions in Android are two straightforward examples of it. In practice, behaviour of deceptive apps can lead to the violation of users' data privacy when the apps are not transparent about complete behaviour specifications to unsuspecting users. 

\subsubsection{Security, Privacy, and Protection}
The terms "security", "privacy",  and "protection" are frequently used in connection with Android and information systems. It is noteworthy that all authors do not use these terms in the same way. This study refers to definitions commonly encountered in computer science and Android research literature.  
The term "security" describes techniques that control who may use or modify the computer or device or the information contained in it. 

The term "privacy" denotes a socially defined ability of an individual (or organization) to determine whether, when, and to whom personal (or organizational) information is to be released. 

The term "protection" is used more specifically to denote mechanisms and techniques that control the access of executing programs to stored information on a device \cite{Saltzer1451869}. 

\subsubsection{Potentially Harmful Applications (PHAs)}
Potentially Harmful Applications are apps that can put user data or devices at risk. Some common types of PHAs are Trojans, spyware, ransomware, adware, phishing apps, click fraud apps, etc.  \cite{SecurityReports} PHAs can also come as pre-installed apps on devices because of fragmented Android ecosystem. Or, they can come via over-the-air updates bundled with some genuine system updates. Sometimes users may install PHAs that can change default settings or disable security features or root devices or weaken existing security policies enforced. One study reveals that PHAs can remain active for an average of 77 days on Google Play and 34 days on third-party app stores \cite{temporalmal277110}. Android system should prompt clear messages about risks involved and alert users to make an informed decision. 

\subsubsection{Unwanted Software (UwS)}
Unwanted software applications are apps that are not classified as malware as yet but are harmful apps to the app ecosystem. Google defines Mobile unwanted software (MUwS) as apps impersonating other apps or apps collecting at least one of these without user consent: device phone number, primary email address on the device, querying other installed apps, and information about third-party accounts associated \cite{SecurityReports}. These deceptive apps are over-privileged and pose a risk to users' data privacy. %
Swajcer et al. \cite{sophos_pua} provided a taxonomy of Potentially Unwanted Applications (PUA) in the context of mobile apps. PUA includes and is not limited to Spyware, Adware, hacking tools, low-reputation apps, etc. 

\subsubsection{Malware}: Malware is a kind of software with malicious intentions, to perform harmful operations without the knowledge of users \cite{URL:google_malware}. This software can steal data or even take privileged control of the device on which it runs. Often, malware writers use stealth techniques to evade the detection from existing protection methods \cite{ParvezCOMST6999911}. 



\subsubsection{Android API, Source and Sink}
With each major release of Android OS, an API level is assigned which denotes an update to a set of API calls available with the new release. Android API calls can be source or sink API. 

Android source APIs are defined as the calls into resource methods that return variable values into the app, e.g., getDeviceId(). It returns the International Mobile Equipment Identity (IMEI) number. The IMEI number is a unique value on every mobile phone. Android sink APIs are defined as the calls into resource methods that accept at least one variable value from the app as one of the parameters, and if a new value is written or an existing value is overwritten on the resource. e.g. for user messaging call to sendTextMessage(). It receives message text body and phone number as parameters, both are variable values.

\subsection{Analysis Mechanisms}
There are two primary mechanisms used by researchers to understand and analyze application behaviour - static and dynamic analysis. Static analysis studies applications by reading and inspecting it - be it human-readable source code or some form of machine-readable binary code. Dynamic analysis studies applications by executing them - be it the execution of test cases or running it in a controlled environment. 

\subsubsection{Static Analysis}
Static Analysis involves the traversal of code for all execution paths to analyze all possible behaviours of the application. This analysis allows seeing the code with possible evidence of hypothetical behaviour that can result in malicious behaviour or access control violation or possible privacy leaks. Static analysis cannot produce a trace, run-time context, or actual instances where such a problem occurs. However, it can suggest that such a problem is likely to arise if available code gets executed. Static detection tools use de-compilation or reverse engineering techniques to analyze code and associated data flow or control flow \cite{ParvezCOMST6999911}, \cite{lili10.1016/j.infsof.2017.04.001}. 

Static Analysis can be performed using automation and at a fairly large scale for thousands or hundreds of thousands of applications. There are techniques through which app developers can evade the detection of behaviour by static analysis using code transformations, code obfuscation, dynamic code loading, etc. Static analysis may not be sufficient in these cases. 

\subsubsection{Dynamic Analysis}
Dynamic analysis involves the actual execution of application code and thereby analyzing and auditing its runtime behaviours. Dynamic Analysis suffers from test coverage problems - enumerated test cases can be executed and execution of all possible paths is not practical. Analyzing a large number of applications at scale using dynamic analysis techniques is neither a straightforward nor a trivial task. 

\subsubsection{Hybrid Analysis}
Both static and dynamic analysis techniques can be employed together to analyze application behaviour. This kind of hybrid analysis helps to overcome two major limitations of static and dynamic approaches - application code coverage and scalability of analysis by and large. One can employ static analysis techniques to detect all possible execution paths of interests and further dynamic analysis techniques can help to prune those combinations and identify more likely scenarios and actual instances where problems remain. Alternatively, using dynamic analysis one can find potential use cases and further using static analysis these can be generalized for analysis at scale. 

\begin{table*}
\renewcommand{\arraystretch}{1.8}
\caption{Mapping between API and permissions - Contributions and Comparison}
\label{tab:contributions_comparison}
\centering
\begin{tabular}{|c|c|c|c|c|c|c|}
\hline
\multirow{2}{*}{\bfseries Related Study} & \multirow{2}{*}{\bfseries Year} & \multicolumn{5}{c|}{\bfseries Proposed solution}\\
\cline{3-7}
& & \bfseries Android Version & \bfseries API Level & \bfseries Source and dataset & \bfseries Extension & \bfseries Contribution \\
\hline
Stowaway, & 2011 & 2.2 & 8 & No source code and & Cannot be & Dynamic analysis based on method invocation \\
Felt et al. \cite{Stowaway10.1145/2046707.2046779} & & & & no dataset is available & extended & (API fuzzing) and permission entry point \\
& & & & & & monitoring, extracted sensitive API list \\
\hline
PScout, & 2012 & 2.2.3-5.1 & 10-22 & Source and dataset & No & Static analysis based on CFG and\\ 
Au et al. \cite{Pscout10.1145/2382196.2382222} & & & & available & & reachability analysis and CHA \\
& & & & & & provides permission and API map\\
\hline
Bartel at al. \cite{bartel6813664} & 2014 & 4.0.1 & 14 & Source and dataset & No & Static analysis based on class\\ 
 &  &  &  & not available &  & hierarchy (CHA) and field-sensitive \\
 &  &  &  &  &  & analysis (Spark) \\
\hline
Axplorer & 2016 & 4.1.1, 4.2.2, & 16, 17, & Source not available, & No & Static analysis using inter-procedural \\ 
Backes et al. \cite{axplorer197193} &  & 4.4.4, 5.1 & 19, 22 & dataset available &  & call graph and permission locality in \\
 &  &  &  &  &  & permission checks \\
\hline
DPSpec & 2017 &  &  & Source and dataset & No & Static analysis based on Android \\ 
Bogdanas et al. \cite{bogdanas2017dperm} &  &  &  & not available &  & SDK Source and code, \\
 &  &  &  &  &  & javadoc annotations \\
\hline
Dypermin & 2018 & 4.1.1, 6.0 & 16, 23 & Source code available,& No & Dynamic analysis using Android\\ 
Lyvas et al. \cite{DyperminLYVAS2018472} &  &  &  & dataset not available &  & SDK source \\
 &  &  &  &  &  & \\
\hline
Arcade & 2018 & 7.1 & 25 & Source not available, & No & Using path-sensitive static analysis \\ 
Yousra et al. \cite{Arcade10.1145/3243734.3243842} &  &  &  & dataset available &  & extracted conditional map with \\
 &  &  &  &  &  & UID checks and user checks \\
\hline

\end{tabular}
\end{table*} 

\begin{table*}
\renewcommand{\arraystretch}{1.8}
\caption{Permission Specifications: The full list of evaluated publications}
\label{tab:evaluated_publications_map}
\centering
\begin{tabular}{c c c l c}
\hline
Year & Venue Type & Venue Name & The title of publication & Reference\\ \hline
2011 & ACM & CCS & Android Permissions Demystified & \cite{Stowaway10.1145/2046707.2046779}\\ \hline
2012 & ACM & CCS & PScout: analyzing the android permission specification & \cite{Pscout10.1145/2382196.2382222}\\ \hline
2014 & IEEE & TSE & Static Analysis for Extracting Permission Checks of a Large Scale & \cite{bartel6813664} \\
 &  &  & Framework: The Challenges and Solutions for Analyzing Android & \\ \hline
2016 & USENIX & USENIX Sec & On Demystifying the Android Application Framework: & \cite{axplorer197193}\\ 
 &  &  & Re-Visiting Android Permission Specification Analysis & \\ \hline
2017 & ArXiv & ArXiv & Dperm: Assisting the migration of android apps to runtime permissions & \cite{bogdanas2017dperm}\\ \hline
2018 & ScienceDirect & Computers & Dypermin: Dynamic permission mining framework for android platform & \cite{DyperminLYVAS2018472}\\ 
 &  & and Security & & \\ \hline
2018 & ACM & CCS & Precise Android API Protection Mapping Derivation and Reasoning & \cite{Arcade10.1145/3243734.3243842} \\ \hline

\end{tabular}
\end{table*} 

\section{Mapping between API calls and permissions}\label{sec:mapping_between_API_calls_and_permissions}

In this section, we describe the seminal research work on generating or extracting permission specifications during the years 2011-2020. Table \ref{tab:contributions_comparison} shows an effective comparison to emphasize the contribution of this study in filling the literature gap in the Android permission-based analysis. 

\subsection{Individual Studies}
Felt et al. \cite{Stowaway10.1145/2046707.2046779} developed a tool called Stowaway to detect over-privilege in apps using a study of over 940 apps and found one-third of them are over-privileged. Felt et al. applied automated testing using Randoop \cite{Randoop10.1145/1297846.1297902} an object-oriented test generator for Java programs to Android 2.2 to find out API calls an app uses and further list out permissions necessary to invoke each Android API call. Because of the testing-based dynamic analysis approach, Stowaway suffers from code coverage problems. Being unable to handle reflective calls, the authors identified Java reflection as an open problem. Felt et al. also described permission enforcement mechanisms and some reasons for unwanted permissions found in apps. This work requires modification to the Android source and some manual intervention during app analysis. There is no source and permission map available from the work of Felt et al. and it cannot scale to recent releases of Android. 

Au et al. \cite{Pscout10.1145/2382196.2382222} proposed the tool PScout to extract permission map based on static analysis of Android source code. PScout performs Java bytecode analysis using Soot framework \cite{soot10.5555/781995.782008}. Au et al. generate a Control Flow Graph (CFG) based on Class Hierarchy Analysis (CHA) and perform a backward reachability traversal of CFG to approximate API calls making a permission check or otherwise. It has been observed that PScout generates more permissions to call mappings due to static references of API call strings. 

Bartel et al. \cite{bartel6813664} use static analysis techniques to extract the permission mapping. They generate a Control Flow Graph based on Class Hierarchy Analysis with the help of Soot \cite{soot10.5555/781995.782008} and Spark \cite{spark10.5555/1765931.1765948} frameworks for Android 4.0.1. Although in their research findings, they mention that analysis produces promising results, they have neither provided the source nor pointed to any experimental/empirical data for verification. The permission map generated by Bartel et al. is not available anymore. 

Backes et al. \cite{axplorer197193} provide the first high-level classification of the Android framework's protected resources using Axplorer. They present a novel mapping that significantly improves on PScout results that lack knowledge about the framework's internals. The authors provide three different mappings, namely SDK, Application Framework, and Content Providers for API level 16 (Android v4.1) to 25 (Android v7.1). Backes et al. use the concept of permission locality to demonstrate that permissions checks inside the framework to protect sensitive operations do not follow the principle of separation of duty. 

Bogdanas et al. \cite{bogdanas2017dperm} proposed DPSpec in which the permission map is generated using several documentation formats - annotations in XML files and Java documentation comments of Android SDK. They claim a better accuracy compared to Axplorer. They provide neither the source code of the DPSpec solution nor the permission map for comparison. 

Lyvas et al. \cite{DyperminLYVAS2018472} proposed Dypermin for compiling Android permission specifications using security exceptions during runtime, the availability of any protected API method, and Java reflection mechanism. Dypermin relies on runtime exceptions raised when any API method is invoked without required permissions, and it does not generate false positive mapping entries.

Yousra et al. \cite{Arcade10.1145/3243734.3243842} proposed ARCADE to demonstrate conditional permission specifications by considering user checks under which the condition API is invoked. Authors take path-sensitive control flow graphs into consideration to derive precise permission mapping that contains user checks, e.g. UID. This mapping cannot be used as-it-is in other analyses, and the authors do not provide the source code. 

\subsection{Challenges of permissions map}
In recent years, research work on generating mappings from permissions to API calls has stagnated. There are a few contributing factors to this state. With the growth of features, the Android source code base has increased manifold, and permissions have increased. Along with Java, Kotlin has made headways into the Android source tree, adding complexity to analyzing framework sources. While Google or Android Open Source Project (AOSP) provides neither execution specifications of apps nor tools to analyze framework sources at scale, the relevant documentation is often incomplete. 

\section{Evolution of Permission Model}\label{sec:evolution_permission_model}
Android security model is a multi-party consent model enforcing that an action should happen if all three stakeholders (device user, app developer, platform, or service provider) consent to it \cite{rene10.1145/3448609}. If any party does not consent, by default, the action should be blocked. 

Permissions are a critical part of security and privacy mechanisms embedded in Android. In general, permissions serve two purposes: enforce protection policy and describe the capabilities of an app. 

\subsection{Requesting and Granting Permissions}
With the evolution of the Android framework, how apps request permissions and how permissions are granted - has also gone through several changes. The following summarizes the general method of requesting and granting permissions; we describe evolutionary changes thereafter: 

  \begin{itemize}
    \item Android runs each app with a distinct system identity by assigning a unique user ID - UID to each app. 
    \item Each app runs in a process sandbox, thus isolated from other applications on the device.
    \item Hence app explicitly requests access to resources and data outside its own sandbox. 
    \item App requests such access by declaring required permissions. 
    \item Permissions can be requested \emph{at the time of installing the app} or \emph{at the time of executing the app (run-time)} later. 
    \item System may grant permissions automatically or prompts the user to approve/reject permissions. 
    \item Permissions are specified in the app manifest file. These are presented to the user in the form of a single page during installation or upgrade of the app or in the form of multiple onscreen dialogue boxes while using a specific functionality, e.g. using a camera or microphone. 
    \item Some permissions are granted without an interface to the user. These are normal permissions as defined by Android and are hidden under a collapsed display. 
    \item In terms of notifying permissions to users, it has been observed that permission details are hidden many clicks away from the place of install option provided to the user. Even an informed and experienced user finds it cumbersome to locate such information before installing the app. 
  \end{itemize}

Permissions are categorized into install-time permissions, runtime permissions, and special permissions \cite{URL:requestingPerms}. Install-time permissions consist of normal permissions and signature permissions. Normal permissions are granted at app installation time. Signature permissions are granted at installation time to an app if the signature of this app matches the signature of the app that defined these permissions. Runtime permissions consist of some permissions with \textit{dangerous} protection level. Some special permissions are used by Android platforms and device manufacturers (OEMs) with their implementation details with \textit{appop} protection level. 

\begin{figure}
    \centering
    {
        \subfigure
        {
            \includegraphics[width=90mm]{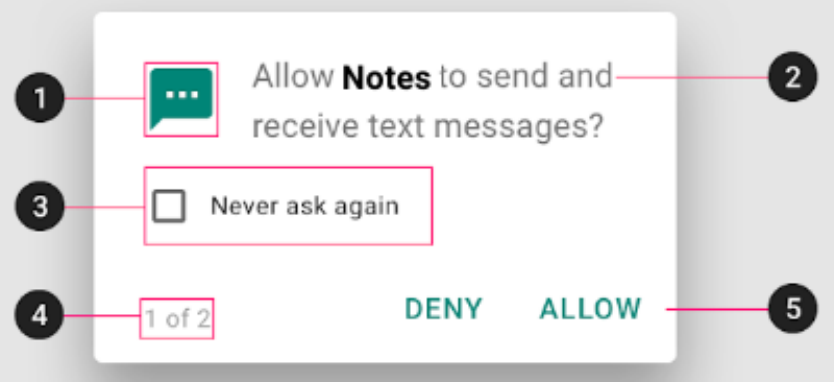}
        }
    }
    \caption{Requesting permission dialog box at runtime}
    \label{fig:permDialog}
\end{figure}

To request permissions at runtime, one or more permission dialog boxes are displayed from Android v6 API level 23. Figure \ref{fig:permDialog} illustrates a sample permission dialog box with labels while requesting permissions at runtime. The labels are summarized below. 

\begin{enumerate}
    \item An icon for permission group, e.g. SMS. 
    \item The name of the app. 
    \item A checkbox for “Never ask again” is displayed when a user denies permission twice.
    \item Count indicator for multiple dialog boxes.
    \item Request-related action for the user - Allow or Deny. 
\end{enumerate}

\begin{table*}
\renewcommand{\arraystretch}{1.8}
\caption{Permission Evolution: The full list of evaluated publications}
\label{tab:evaluated_publications_evol}
\centering
\begin{tabular}{c c c l c}
\hline
Year & Venue Type & Venue Name & The title of publication & Reference \\ \hline
2009 & IEEE & S\&P & Understanding Android Security & \cite{Enck4768655}\\ \hline
2011 & USENIX & USENIX Sec & A Study of Android Application Security & \cite{EnckDed10.5555/2028067.2028088}\\ \hline
2012 & ACM & ACSAC & Permission Evolution in the Android Ecosystem & \cite{Wei2012evolution} \\ \hline
2016 & Springer & RAID & Small Changes, Big Changes: An Updated View on the & \cite{yuri10.1007/978-3-319-45719-2}\\ 
 &  &  & Android Permission System & \\ \hline
2020 & ACM & MSR & Automatically Granted Permissions in Android Apps: An Empirical & \cite{Autogranted10.1145/3379597.3387469}\\ 
 &  &  & Study on Their Prevalence and on the Potential Threats for Privacy &  \\ \hline
\end{tabular}
\end{table*} 

\subsection{Evolutionary changes}

Enck et al. \cite{Enck4768655} first studied Android's security model and uncovered complexity in the very early days of Android. Enck et al. \cite{EnckKirin10.1145/1653662.1653691} proposed Kirin to use template-based security rules to detect dangerous functionality implemented in apps. In another work, Enck et al. \cite{EnckDed10.5555/2028067.2028088} presented a Ded compiler to recover source code from Android applications. They uncovered the usage of phone identifiers for advertising and analytical networks. Yuri et al. studied permission-related changes in Android API level 23 or before \cite{yuri10.1007/978-3-319-45719-2}. Jeon et al. \cite{finegrained10.1145/2381934.2381938} proposed to detect fine-grained permissions required using RefineDroid and enforce these permissions in modified app binary. Further, we present how some of the major changes have been introduced in Android. 


\begin{figure*}[!ht]
\centering
\includegraphics[width=160mm]{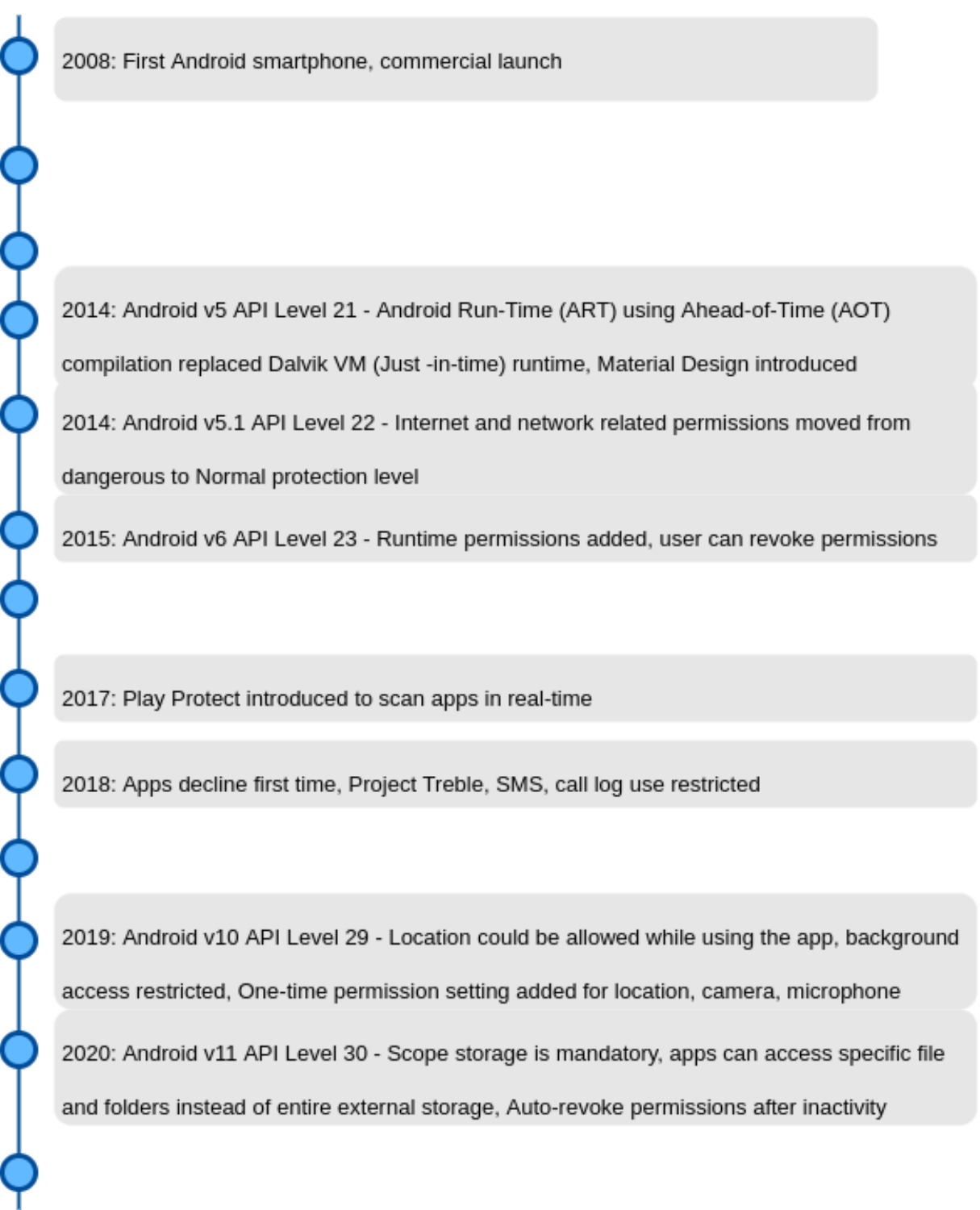}
\caption{Timeline for evolution in Android in the past decade.} 
\label{fig:andytimeline}
\end{figure*}

\subsubsection{Grant all permissions at install time}
Prior to Android API level 23 (Android 6.0 Marshmallow), when the user wanted to install an app, the installer used to show all permissions-related text. For the installer to proceed, it was mandatory to consent and grant all permissions required by the app. There was no choice or discretion with the user to selectively not grant specific permissions at installation time itself. Moreover, there was no provision to revoke granted permissions. This all-or-none mandatory behaviour could trick users into vacuously granting all permissions without going into details of the permissions being asked. Up to API level 22, all permissions requested in the app manifest were granted. 

\subsubsection{Run-time permissions} 
From Android 6.0 API level 23, permissions are divided into install-time (critical for the app to work) and run-time permissions based on the time they are requested first to the user for consent \cite{devandy_6.0behaviour}. Normal and signature (with few exceptions) permissions are permanently granted when the user installs the app. Dangerous permissions are now granted at run-time, and the user can revoke them anytime. After much criticized install-time permissions, this was a significant change from Android 6.0, which allowed users to revoke permissions selectively at the group level. In general, specific permissions are granted to the app by default, and for some permissions, the user is presented with a prompt to allow or deny permission. However, users can revoke one or more permission using app configuration settings. API level 23 significantly changed in that not all permissions requested in the app manifest are granted at install time. This change was due to the realization that not all users are sufficiently equipped to make such a decision at install time \cite{Wijesekara2015android}. 

\subsubsection{SELinux}
Android uses Security-Enhanced Linux - SELinux to enforce mandatory access control (MAC) over all processes. This includes processes with root or superuser privileges. SELinux works on the principle of default denial, wherein anything not explicitly allowed by access control policy is considered as denied. 

\subsubsection{Storage permissions}
All Android apps had access to read content stored on external SD card storage at app installation up to API level 18. From API level 19 and later, two storage permissions are enforced - READ\_EXTERNAL\_STORAGE and WRITE\_EXTERNAL\_STORAGE. Also, these permissions are not required to read or write files in the application-specific directories on external storage. Further, any app that declares the WRITE\_EXTERNAL\_STORAGE permission is implicitly granted READ\_EXTERNAL\_STORAGE permission \cite{platform-manifest}.

\subsubsection{MAC address randomization}
Android added MAC address randomization in v8.0 \cite{randmacBadhwar2021} to prevent listeners and traffic sniffers to create a device MAC-based profile. This change was aimed at limiting the tracking capabilities of apps and increasing user privacy \cite{macrandom}. However, user tracking networks and targeted advertising agencies use other identifiers (IMSI, Ad ID or ANDROID\_ID, Google Service Framework (GSF) ID, etc.) to uniquely determine users to a large extent or at an abstract group level. However, in Android v10 API level 29, for enterprise device usage, device owner apps can retrieve MAC using getWifiMacAddress() API \cite{devandy_10.0privacy}, thereby allowing enterprise device management remotely. 


\subsubsection{Context dependent permission only while using the app}
In 2019, Android v10 API level 29 added a non-binary context-dependent permission setting for accessing the location data of the device. In addition to \textit{Allow all the time} and \textit{Deny}, a third setting was added to set location access to \textit{Allow only while using the app}. This was aimed at bringing location transparency and giving more control to restrict location data usage when it is not needed \cite{URL:locationcontrol}. Interestingly, the third setting grants the location permission when the app is active in the foreground, thereby restricting access to location data stealthily in the background. %
Starting Android v11 API level 30, Camera and Microphone permissions were also subjected to this third setting to \textit{Allow only while using the app}. This move blocks apps from accessing the Camera or Microphone in the background without active notice of the user \cite{rene10.1145/3448609}. Nauman et al. \cite{ApexConditionAllow10.1145/1755688.1755732} had earlier presented an extension allowing permissions conditionally instead of allowing them all the time. 

\subsubsection{One-time permissions} 
From Android v10 API level 29, when apps request permission for location, microphone, or camera, the user can opt for “Only this time”. This provision grants a temporary one-time permission for the app, thus limiting the app’s ability to access sensitive data each time the app is started. 

\subsubsection{Auto-reset permissions} 
If the user has not used an app for a few months on Android 11, the system auto-resets the sensitive permissions. However, the duration after which this change comes into effect is not clear. It has been observed that while reviewing these permissions under the settings of apps, some of these reset permissions are added back to apps. 

\subsubsection{Scoped storage} 
Android API level 29 introduced the concept of scoped storage which was made mandatory from API level 30 onward. Earlier apps could access entire external storage instead of specific files and folders. If scoped storage is enabled, apps can read or write specific files and folders instead of entire external storage. This scoped storage presents a significant improvement to the privacy of users' content on the device. 


\subsection{Changing permission groups}
Along with permissions, permissions groups have evolved. Dangerous permissions are divided into groups to access sensitive resources. There were nine permission groups till Android v8 API level 27 - Calendar, Camera, Contacts, Location, Microphone, Phone, Sensors, SMS, and Storage. In Android v9 API level 28, Call Log was added to facilitate reading and editing call logs by restricted apps. Further, in Android v10 API level 29, Activity Recognition has been added. Bundling write or edit permissions along with read permission for specific resources together poses a risk to the granularity of granting permissions.

\section{Checking Permissions}\label{sec:checking_permissions}
In this section, we survey how permissions are checked, and we characterize permissions. 

\begin{table*}
\renewcommand{\arraystretch}{1.8}
\caption{Permission Checking: The full list of evaluated publications}
\label{tab:evaluated_publications_check}
\centering
\begin{tabular}{c c c l c}
\hline
Year & Venue Type & Venue Name & The title of publication & Reference \\ \hline
2011 & ACM & CCS & Android Permissions Demystified & \cite{Stowaway10.1145/2046707.2046779}\\ \hline
2014 & IEEE & TSE & Static Analysis for Extracting Permission Checks of a Large Scale & \cite{bartel6813664} \\
 &  &  & Framework: The Challenges and Solutions for Analyzing Android & \\ \hline
2022 & IEEE & TSE & Runtime Permission Issues in Android Apps: Taxonomy, Practices,  & \cite{ARPissues9705152} \\ 
 &  &  & and Ways Forward & \\ \hline

2022 & ACM & ICSE & APER: Evolution-Aware Runtime Permission Misuse Detection & \cite{APERarxiv.2201.12542} \\ \hline
\end{tabular}
\end{table*} 

\subsection{Permissions Checks}
The purpose of permission checks is to determine if the app holds permission to execute an API call or access privileged resources. Permission checks can occur at two places: (a) in the app binary when the app uses Android API and (b) in the system process where the actual API is implemented. We observed that permission checks are spread between the Java API layer and Native C/C++ libraries and a few permissions are enforced at the Linux kernel level. 

Android app should declare permission in the app manifest file (AndroidManifest.xml) using the <uses-permission> XML markup. From Android API level 23 onward, the normal permissions are granted vacuously, without any prompt to the user. The dangerous permissions should be requested explicitly to avoid any runtime permission (ARP) issues \cite{ARPissues9705152}. 

\begin{figure}
    \centering
    {
        \subfigure
        {
            \includegraphics[width=90mm, height=120mm]{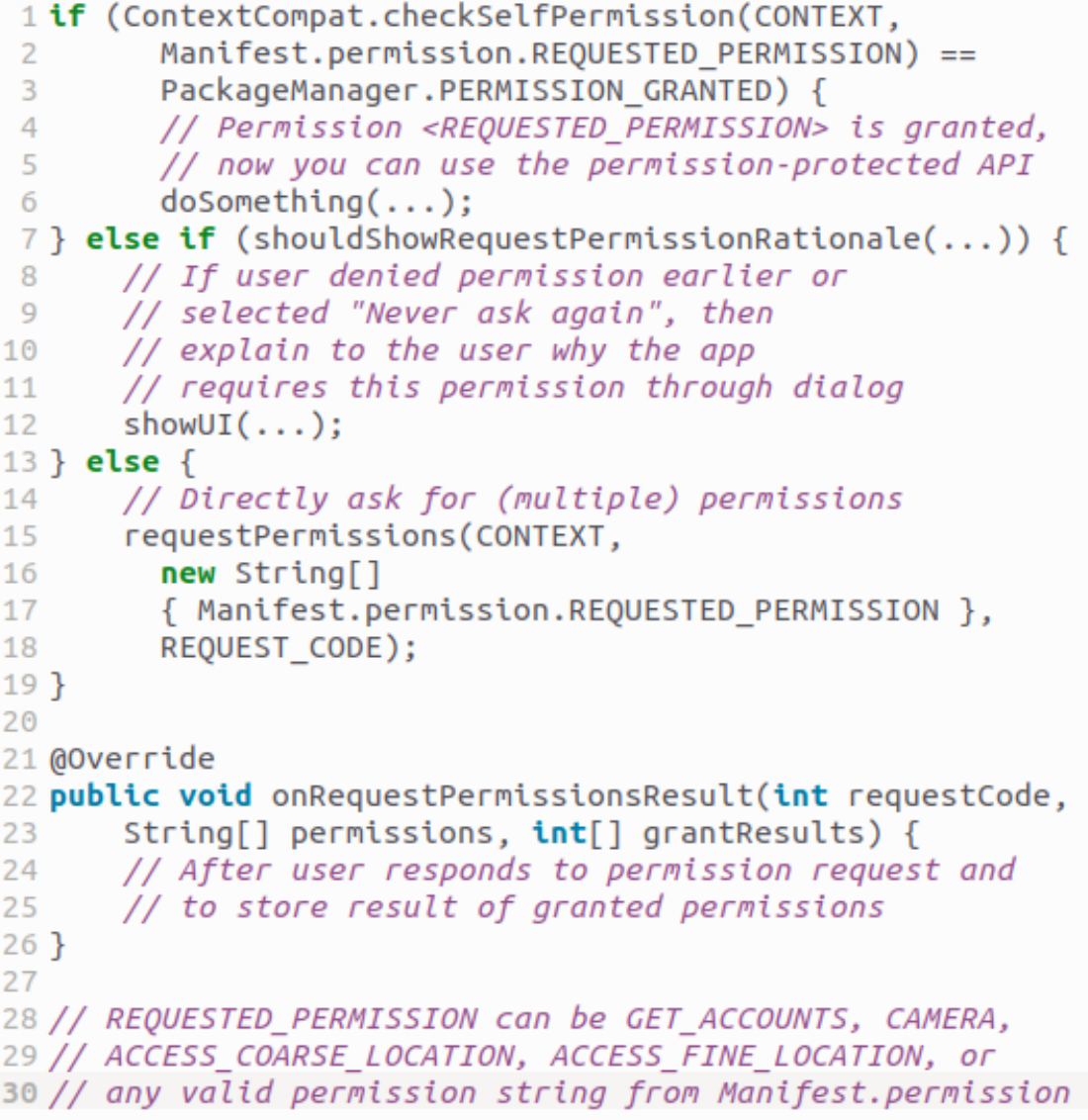}
        }
    }
    \caption{Sample code to check, request and explain permissions}
    \label{fig:checkperms}
\end{figure}

As a best practice, before calling permission-protected APIs, the app should check if permission has been granted to the app by calling checkSelfPermission() API \cite{URL:requestingPerms}. The app should exercise caution to request any dangerous permission at runtime by calling requestPermissions() API. This will alert user with a pop-up dialog box to grant or deny permission or even block any subsequent requests ("Never ask again" option). The app can store results of granted permissions using onRequestPermissionsResult() API and respond to the user \cite{URL:requestingPerms}. A partial code snippet is shown in Figure \ref{fig:checkperms} to illustrate these major permission management APIs. 

Android operating system was envisaged as an open alternative to closed smartphone OS platforms. However, upon closer inspection, we find that there are closed elements of the Android platform. There are closed portions of Google's Android code base (such as device drivers). Users are cut off from accessing some core elements of the Android such as the minified kernel and system daemons employing software and policy mechanisms. The control lies with Google, cellular carriers, and smartphone manufacturers (OEMs). Due to this restriction embedded in the Android ecosystem, specific implementation details related to permission checks are not widely available. 

\subsection{Characterizing Permissions}

Internet and other network-related permissions were moved to normal protection level in API level 22 \cite{yuri10.1007/978-3-319-45719-2}. While it is debatable whether moving the internet to a normal protection level has done any good, it is evident that apps do not explicitly seek users' attention to access the internet. The same holds for any third-party libraries used by apps. 

Read SMS permission is considered intrusive by many users and developers because an app acquiring this permission can read all SMSes on the device. This permission was restricted after Play Store started asking for a six-page documentation on why this permission was required as part of Project Strobe \cite{URL:projectstrobe}. As a result, the number of apps with access to SMS and Call Log permissions reduced by a surprising 98\% within a year \cite{URL:projectstrobeupdate}. 



Location and background location: There are two permissions related to location that the app can request - ACCESS\_COURSE\_LOCATION and ACCESS\_FINE\_LOCATION.  \\

A special permission SYSTEM\_ALERT\_WINDOW was added in Android 6.0 (Marshmallow) to create overlay windows displayed on top of screens of all other apps. A few system-level apps should use this permission to interact with users. An app for API level 23 or higher must explicitly request this permission for user's consent by sending an explicit intent with action ACTION\_MANAGE\_OVERLAY\_PERMISSION. The app should check permission authorization by calling android.provider.Settings.canDrawOverlays()  \cite{platform-manifest}. An early release of Android 6.0 granted this dangerous permission to all apps. Fratantonio et al. demonstrated several UI attacks using only two permissions - SYSTEM\_ALERT\_WINDOW and BIND\_ACCESSIBILITY\_SERVICE in the Cloak and Dagger manner. 

Another special permission WRITE\_SETTINGS allows an app to read or write the system settings. An app for API level 23 or higher must explicitly request this permission for the user's consent by sending an intent with action ACTION\_MANAGE\_WRITE\_SETTINGS. The app should obtain permission authorization by calling android.provider.Settings.System.canWrite()  \cite{platform-manifest}.

\subsection{3rd Party library usage}
\begin{figure*}
    \centering
    {
        \subfigure
        {
            \includegraphics[width=160mm,height=90mm]{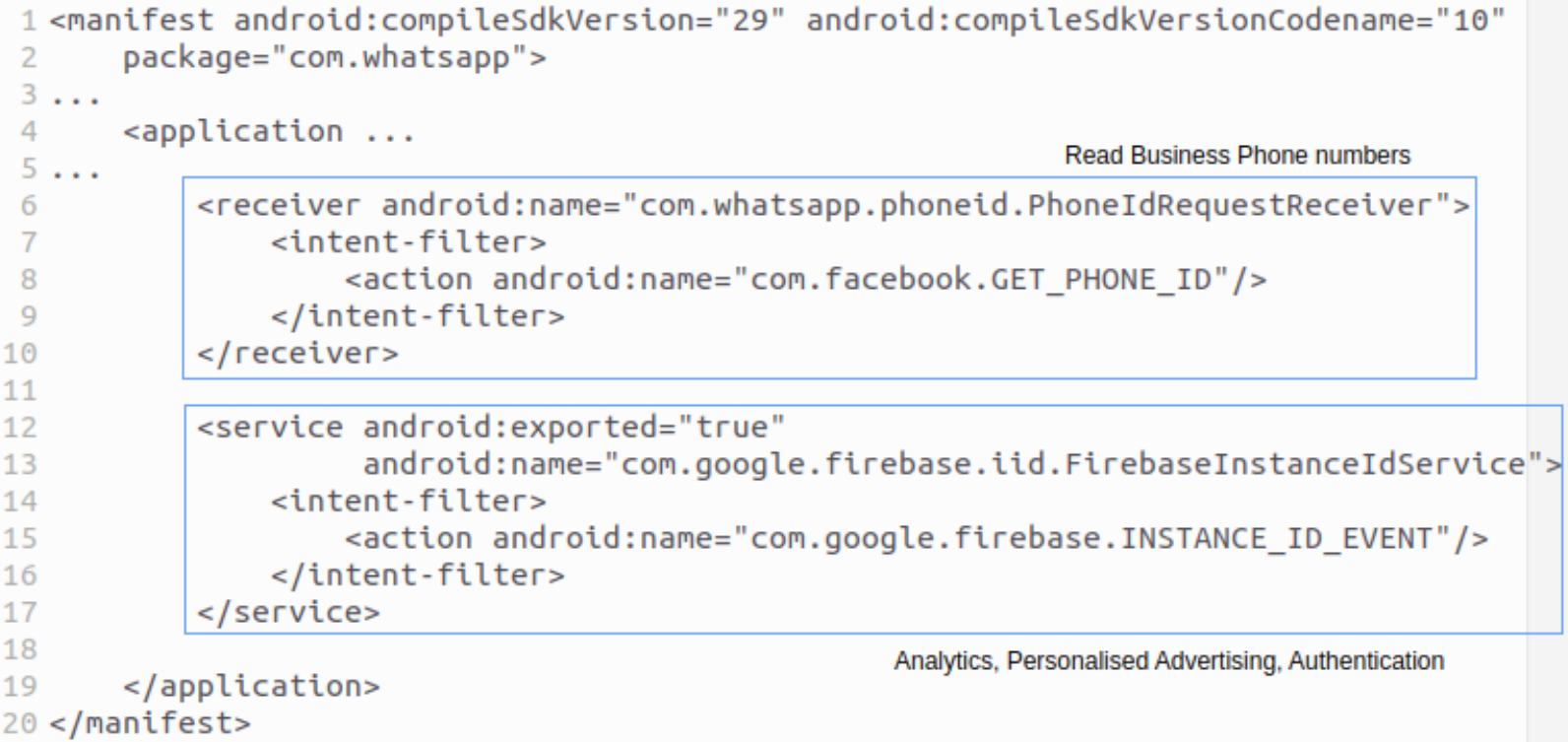}
        }
    }
    \caption{App manifest showing 3rd party library used for analytics and advertising}
    \label{fig:sample_app_manifest_ad}
\end{figure*}

Free apps in the Android ecosystem are supported and monetized by advertising networks, trackers, and data brokers. These 3rd party organizations supply software libraries to offer integration for ad delivery and tracking users without user consent \cite{track-razaghpanah2018}. It is noteworthy that advertising and analytical libraries and app code run with the same access privileges and thereby have access to the same permissions as the app itself. Thus, these 3rd party libraries can share device identifiers and user-specific sensitive information with data brokers for selling ads. This poses a security and privacy risk to data stored and submitted by the end users. Figure \ref{fig:sample_app_manifest_ad} shows a sample app manifest with Firebase library reference \cite{URL:firebase_uc}. 

AppCensus project helped to learn the privacy cost of free apps. \cite{URL:appcensus}. A study on one million websites \cite{10.1145/2976749.2978313} for web privacy measurement revealed that every 3rd party service provider is potentially a tracker. With a mobile-first push, apps can similarly track user activity and even with a broader reach. 


\subsection{Reasons for Permission Gap}\label{subsec:reasons_of_permission_gap}
From the studies mentioned in this paper, it is evident that apps do declare more permissions than required. In this section, we summarize a causal analysis of the permission gap that arises due to over-privileged Android apps binaries: 

\begin{table*}
\renewcommand{\arraystretch}{1.8}
\caption{Permission Over-privilege: The full list of evaluated publications}
\label{tab:evaluated_publications_over-priv}
\centering
\begin{tabular}{ c c c l c }
\hline
Year & Venue Type & Venue Name & The title of publication & Reference \\ \hline
2011 & IEEE & W2SP & Curbing Android Permissions Creep & \cite{vidas11curbing} \\ \hline
2011 & ACM & CCS & Android Permissions Demystified & \cite{Stowaway10.1145/2046707.2046779} \\ \hline
2013 & USENIX & USENIX & WHYPER: Towards Automating Risk Assessment of & \cite{RPandita2013whyper} \\ 
 &  & Security & Mobile Applications & \\ \hline
2014 & ACM & CCS & AutoCog: Measuring the Description-to-permission Fidelity & \cite{Zhengyang2014autocog} \\ 
 &  &  & in Android Applications & \\ \hline

2021 & ScienceDirect & Computer \&  & MapperDroid: Verifying app capabilities from description to & \cite{MapperDroidSOLANKI2021102493} \\ 
 &  & Security & permissions and API calls & \\ \hline

\end{tabular}
\end{table*} 

\begin{enumerate}[label=\alph*)]
    \item Network Permissions requested together or in pairs: It is often seen that apps request Internet, Wi-Fi, and network access-related permissions together while the app does not even use them. This results due to the misconception that every app requires internet and network-related permissions always. It is possible that app needs one of these permissions while app developers add them all. Of the applications that unnecessarily request WiFi or network permission, a significantly lower percentage of apps legitimately require network or WiFi permission, respectively \cite{Stowaway10.1145/2046707.2046779}. 

    \item Availability of poor or incomplete documentation: App developers often refer to the Android developer reference manual \footnotemark[1] for API-related documentation. It is known that API documentation is not complete or up-to-date, and there are hidden APIs \cite{URL:anggrayudi/hidden-2020}. It is also a reason that developers who are not very experienced may not find relevant information and may end up using permission that they could have avoided in the first place with better help available. 
    \footnotetext[1]{https://developer.android.com/docs}

    \item Deprecated permissions in use: In the Android app manifest file, developers specify the minimum and maximum API levels they target. Subsequent app updates can change this information in the manifest header. While app updates, it is possible that the developer did not pay attention to removing deprecated permissions as Android API levels introduce new behaviour changes. Since app development tools or IDEs do not enforce this removal, deprecated permissions also find their way into released apps. Users may not even be informed of dormant permissions usage while they are invoked by any upgrade \cite{Sellwood2013SleepingAT}. 
    
    \item Active Android API level spread: As of early 2021, Android API level 30 (version 11) is the latest stable release available. While there exists a non-negligible share of active Android devices using API level 19-23 (version 4.4-6.0) released way back in 2013 \cite{URL:StatistaAndyPlatforms}. It is noteworthy that from 2013-2015 there were five to six API levels co-existed in active Android devices, which doubled in numbers during 2018-2020. This vast spread of active API levels forces app developers to include a wide range of minimum and maximum API levels to support backward compatibility for the same app binaries, which results in poor or misconfigured applications for newer versions of apps. 
    
    \item Cost of correcting permission over-declaration: Developers are expected to release apps with more and more features in less and less time. They may  over-declaration permissions due to the above-mentioned reasons. The cost of correcting over-declaration is high, i.e. developers will require to investigate the usage of individual permissions and update the manifest. If the Android SDK framework or development tools can alert developers about this over-declaration, they can fix it at a lower cost \cite{short10.1145/2046614.2046626}. 
\end{enumerate}

\section{Potential Future Directions}\label{sec:future_directions}
We have covered some seminal work on Android research focused on permissions in the past 10-12 years (2010-2022). Based on this survey, several insights are drawn and are listed below. 

\subsection{Permissions map for recent Android versions} 
This study recommends the improvement in the availability of mapping between API calls and permissions for Android versions that have come in the past 5 years, specifically Android version 8 or API level 26. Due to the complexity of the Android framework source and the lack of studies on Android internal framework details, research work has stagnated on this front. These permission maps will facilitate research on over-privileged or misconfigured applications, deceptive apps, and malware detection. 

\subsection{Availability of datasets for evaluation} 
In the analyzed studies, the usage of older datasets is predominant, be it malware applications or app metadata-related datasets \cite{deeplearn9211502}. This forces authors of new studies to utilize these older datasets for comparison, thereby raising questions on the relevance in the period of 2018-2022. We observe that none of the studies mentioned in Table \ref{tab:contributions_comparison} can be extended and reproduced for recent Android versions. It is recommended that new studies provide sufficient materials to reproduce work and provide datasets to evaluate the approaches and draw fair comparisons. 

\subsection{Separation of third-party library related permissions}
Android runtime permissions issues may arise if there are conflicts between the configuration files of apps and third-party libraries. These third-party libraries may use permission-protected APIs without properly checking and requesting only required permissions \cite{ARPissues9705152}. Presently, app developers and users are not made aware of specific permissions being required by third-party libraries. Unity framework v5.1 to build apps was adding internet and reading phone status for every app while developers did not intend it. A diabetes app Glucosio was found to have record audio permission due to a third-party library InstaBug in use \cite{URL:GlucosioRecordAudio}. If app developers and end users are provided with information on such permission usage separately, misconfigured or deceptive apps can be detected early. This is applicable for both fresh installations and any subsequent updates of apps. 

\subsection{Custom permissions by 3rd-party apps}
Tuncay et al. \cite{cusper2018tuncay} raised concerns of custom permissions. Authors showed that provisions to regulate access to app components can be misused to gain unauthorized access. 

\subsection{Maximum number of active Android API levels}
As we noted in section \ref{subsec:reasons_of_permission_gap}, there exists a large number of active Android API levels at the same time. If the maximum number of active API levels can be restricted to four or fewer, this will force device manufacturers (OEMs) to upgrade Android major releases faster than in practice. This, in turn, will force app developers to stick closer to and leverage several enhancements happening in successive OS releases. Potentially this initiative can reduce Android attack surface as device software gets upgraded faster than it has been happening as of early 2021. 

\subsection{Granularity of granting permissions}
As of API level 30 and way back when run-time permissions were introduced from API level 23, dangerous permissions are granted at the permission group level. Due to this, apps have access to other permissions in the same group unnecessarily. e.g. app having READ\_PHONE\_STATE permission also has access to CALL\_PHONE and other permissions in the phone group. Paolo et al. show in a recent study \cite{Autogranted10.1145/3379597.3387469} that 56\% of apps use automatically granted permissions, despite their application description not explaining the purposes. It is strongly recommended that permissions should be granted at a finer level than the present permission group to avoid abuse of automatically granted permissions. 

\subsection{Looking beyond one app at a time}
In documented studies, we often find static or dynamic analysis techniques looking at one app at a time and, at times, ignoring other apps on the device by the same developer or different developers. It is possible that one app alone may not ex-filtrate sensitive information, and it may do so with the help of permissions obtained by one or more other apps on the device. 

\subsection{A standard UI for requesting permissions} 
A baseline standard for the User Interface of requesting permission is required. Such a standard interface should clarify why permission or a set of permissions is required. The app should explicitly specify what information will be collected and how collected information will be used. The app should not ask for unnecessary permissions in advance that may be required in a future update. Instead, the user should be prompted in the future to grant those permissions for newly introduced functionality \cite{URL:material-andyperms}.

\section{Conclusion}\label{sec:conclusion}
Several studies over a decade have reported challenges in the implementation of the permission model in Android. If the permission model is not understood appropriately by app developers and end users, it can lead to poor decisions posing a risk to the security and privacy of users' data on the device. This study presents three important dimensions related to Android security research, thereby filling a gap in systematized knowledge of these topics. The study scrutinizes research on permission specifications and suggests the need for such a precise specification for newer Android versions. Knowledge and availability of permission map can impact future studies to detect deceptive apps and Android malware. 

We present the evolution of permissions and seminal contributions to study the challenges in the permission model. The study further builds knowledge on how permissions are checked and granted. This work also recommends future directions based on several studies analyzed. The analysis concludes that studies are scarce in this area regarding analyzing modern-day apps and re-usable datasets for evaluating deceptive app analysis. We also suggest short bibliographies in the form of tables to direct the readers to the most useful, up-to-date, and significant publications available.


\appendices



\ifCLASSOPTIONcaptionsoff
  \newpage
\fi

\bibliographystyle{IEEEtran}
\bibliography{article}

\end{document}